\documentclass{vgtc}                          

\ifpdf
  \pdfoutput=1\relax                   
  \usepackage{graphicx}                
  \DeclareGraphicsExtensions{.pdf,.png,.jpg,.jpeg} 
\else
  \usepackage{graphicx}                
  \DeclareGraphicsExtensions{.eps}     
\fi%

\graphicspath{{figures/}{pictures/}{images/}{./}} 

\usepackage{microtype}                 
\PassOptionsToPackage{warn}{textcomp}  
\usepackage{textcomp}                  
\usepackage{mathptmx}                  
\usepackage{times}                     
\usepackage{cite}                      
\usepackage{tabu}                      
\usepackage{booktabs}                  

\onlineid{0}

\vgtccategory{Research}

\vgtcinsertpkg

\title{Design Judgment in Data Visualization Practice}

\author{Paul Parsons\thanks{e-mail: parsonsp@purdue.edu} %
\and Colin M. Gray\thanks{e-mail: gray42@purdue.edu} %
\and Ali Baigelenov\thanks{e-mail: abaigele@purdue.edu} %
\and Ian Carr\thanks{e-mail: carr80@purdue.edu}} %
\affiliation{\scriptsize Purdue University}

\abstract{Data visualization is becoming an increasingly popular field of design practice. Although many studies have highlighted the knowledge required for effective data visualization design, their focus has largely been on formal knowledge and logical decision-making processes that can be abstracted and codified. Less attention has been paid to the more situated and personal ways of knowing that are prevalent in all design activity. In this study, we conducted semi-structured interviews with data visualization practitioners during which they were asked to describe the practical and situated aspects of their design processes. Using a philosophical framework of design judgment from Nelson and Stolterman \cite{nelson_design_2012}, we analyzed the transcripts to describe the volume and complex layering of design judgments that are used by data visualization practitioners as they describe and interrogate their work. We identify aspects of data visualization practice that require further investigation beyond notions of rational, model- or principle-directed decision-making processes.%
} 

\CCScatlist{
  \CCScatTwelve{Human-centered computing}{Visu\-al\-iza\-tion}{}{};
}

\begin{document}


\firstsection{Introduction}

\maketitle


One of the goals of visualization research is to influence design practice. This aim is becoming increasingly relevant, as data visualization as a profession is rapidly growing in popularity \cite{bigelow_reflections_2014,parsons_what_2020}. To effectively influence design practice, it is necessary to understand the ways in which practitioners structure their work in relation to the complexity they face in real-world settings \cite{Stolterman2008}. While research focusing on tools and issues that practitioners face is gaining traction in InfoVis \cite{bigelow_reflections_2014,bigelow2016iterating,mendez2017bottom,hoffswell2020techniques,walny2019data,parsons_what_2020}, only a small number of studies have been grounded in real-world design practice (e.g., \cite{bigelow_reflections_2014,hoffswell2020techniques,walny2019data,parsons_what_2020,Parsons_chartjunk_2020}). 

Although recent work has highlighted the messy, complex, and situated nature of visualization design\cite{mckenna_design_2014,McCurdy2016,meyer_criteria_2020} and the challenges in evaluating visualization authoring systems \cite{ren2018reflecting,satyanarayan2019critical}, these efforts have focused on visualization \textit{researchers} rather than \textit{practitioners}. Unlike other design-oriented fields, InfoVis lacks a recognizable expansion of scholarly inquiry into design practice as an activity in its own right---distinct from research activity---in ways that can account for the complexity of everyday design situations.

Multiple design-related fields have undergone a turn to practice as a means of informing research, with a recognition that barriers to knowledge production and use have arisen between the academic and practitioner communities. The role of practice-led research is relevant both in emergent design fields such as interaction design and instructional design \cite{boling_core_2017,kuutti_turn_2014}, and in more traditional domains such as architecture and studio art. We position this paper as a means of motivating further work that is practice-focused and practice-led, drawing inspiration from these design fields in an effort to better understand the ways in which practitioners design and the knowledge that can inform future visualization design research.

In this work, we focus on how designers think and make decisions during their design process. Although design decision and process models have previously been proposed in InfoVis (e.g., \cite{munzner_nested_2009,meyer_nested_2015,sedlmair_design_2012,mckenna_design_2014}), the limitations of such models in describing real-world design practice have been noted by design scholars in other fields. \cite{goodman_understanding_2011,gray_judgment_2015,Stolterman2008,lawson2006designers,Cross_1982}. These scholars highlight the personal and situated factors of design practice that are essential for making decisions and moving through a design process---factors that tend to resist modeling and codification. 

We use a philosophical framework from Nelson and Stolterman \cite{nelson_design_2012,Vickers1984-gc} to guide our inquiry into the ways that data visualization practitioners make decisions during their design work. We build upon decades-old interest in \textit{phronesis}---or the practical knowledge that underlies and supports the work of professionals \cite{dunne_professional_1999,Shotter2014-ev}---and investigate how \textit{design judgment}, a particular aspect of that professional engagement, aids a designer in confronting the complexity of real-world design situations and guides decision making \cite{nelson_design_2012}. This work builds upon prior investigation of design judgments in instructional design, where Boling and colleagues described the complex and layered nature of judgments in informing design practice \cite{boling_core_2017,gray_judgment_2015} and design education \cite{Demiral-Uzan2015-al}.

The contribution of this paper is two-fold: First, we identify and describe the volume of design judgments that are used by data visualization practitioners as they describe and interrogate their work, laying the foundation for future study of the complexity of design practices in pedagogy and industry contexts. Second, we characterize the complex layering of judgments both within discrete design acts and over time, demonstrating aspects of data visualization practices that require further investigation beyond notions of rational, model- or principle-directed decision making processes.

\section{Understanding Design Practice}

\subsection{Design Models in InfoVis}
Early InfoVis research was largely tool- and technique-driven, with little emphasis placed on the design process and the role of the designer in that process. Over roughly the past decade, InfoVis researchers have embraced a more expansive view of design, turning attention to both the process designers go through and the decisions they make while designing. Often drawn from personal experience in designing visualization systems, researchers have proposed various models for visualization design (e.g., \cite{chi_taxonomy_2000,munzner_nested_2009, meyer_nested_2015, mckenna_design_2014, McCurdy2016}). These models have especially focused on process (e.g.,  \cite{sedlmair_design_2012}) and decisions (e.g., \cite{munzner_nested_2009, meyer_nested_2015, mckenna_design_2014}) that designers purportedly make while designing.

While design models have been influential in the research community, it is unclear how well they characterize and support design practice, especially in non-research settings where practitioners may ground their work and process in less theoretical and more pragmatic ways. Furthermore, as the development of these models did not involve the study of design \textit{practitioners}, it is unclear how ecologically valid claims of process and decision-making might be ``in the wild'' of practice. In particular, extant investigations have not accounted for the personal, situated aspects of design that significantly influence designers' thinking and are inseparable from the designer. The personal, situated nature of design activity requires a more expansive vocabulary that goes beyond formal decision making to include \textit{judgments} that are undertaken at every stage of visualization design in opportunistic and personal ways. 

\subsection{Decision Making and Complexity}
Rational decision-making processes are often viewed as the standard for dealing with complex situations. However, the study of real-world decision-making has shown that rational processes are the exception rather than the rule when people are faced with uncertainty in complex situations \cite{klein2017sources}. Formal models offer neat and clear descriptions of how people make decisions, yet they do so by neglecting the contextual aspects of decision making that are not easy to model. Empirical investigations show that people instead rely on patterns of experience, making judgments that do not look much like formal decisions at all \cite{beach2017classical}. 

We use the concept of \textit{design judgment}---a particular form of judgment---to investigate how data visualization practitioners engage with design complexity, recognizing that design situations are messy, complex, and situated in the real-world. In this context, designers do not stand apart from the design process---they participate in it, shape it, and mold it using their lived experience and a range of knowledge types. Thus, particular judgments are subjectively positioned, not replicable, and generally resist codification. This subjectivity of practice links to broader understandings of decision-making, such as the role of personal experience \cite{klein2017sources} and other personal commitments in even the most ``objective'' situations \cite{polanyi_tacit_1966}. Although judgments do not look like `rational' decisions, they are not irrational or mere opinion; judgment relies on familiarity and prototypicality, in which patterns of experience are drawn from in a contextually bound manner that resists codification or objectivation \cite{klein2017sources}. It is this space in which we seek to make a contribution in the empirical work that follows.


\section{Method}
As part of a broader effort to investigate aspects of design complexity in data visualization practice, we interviewed practitioners and asked them questions about their design practice. Recruiting was done via social media, the DataVis Society’s Slack workspace, the InfoVis email list, and our personal networks. To mitigate sampling bias, we also searched widely for professionals and agencies, ultimately contacting more than 200 individuals and more than 30 agencies. 

Interviews were semi-structured and were conducted remotely via videoconferencing. For this paper, we selected 10 of the 20 transcripts to analyze, aiming for diversity across the self-reported characteristics (see Table \ref{tab:participants}). We selected one section of the transcripts in which we asked participants to describe: (1) their typical design process at a high level and (2) how they assessed their progress, including determining if they were on the right track and making decisions about what to do next in their process. Depending on the answers given, we asked a number of follow-up questions regarding how participants started their design process, what kinds of considerations they made, whether their process was more structured or unstructured,  and how they would make specific decisions (e.g., choosing chart types or visual encoding channels). This section accounted for roughly 15 minutes near the beginning of the 60-75 minutes of each interview. The transcripts were deductively coded using Nelson and Stolterman's framework on design judgment \cite{nelson_design_2012}. We operationalized the judgment types described in this framework based on prior work by one of the authors in another design discipline \cite{gray_judgment_2015} and used these types as \textit{a priori} codes in our top-down thematic analysis \cite{braun2006using}. Table \ref{tab:judgments} lists these design judgments and their definitions. All judgment types were coded non-exclusively by two or more researchers, with the goal of reaching consensus and full agreement. We regularly discussed our code application to assure consistency of coding and a shared agreement regarding the meaning of each judgment type.

\section{Findings}

\begin{table}
\begin{tabular}{ p{0.05\columnwidth} | p{0.42\columnwidth} | p{0.09\columnwidth} | p{0.13\columnwidth} | p{0.06\columnwidth} }
\textbf{ID} & \textbf{Job Title} & \textbf{Exp. (yrs)} & \textbf{Highest Degree} & \textbf{G} \\ 
\hline
P1 & Sr. UX Design Lead & 8-10 & D & M \\
P2 & Graphics Editor & 8-10 & B & F \\
P3 & Data Communicator & 2-4 & M & M \\
P4 & Sr. DataVis Dev & $>$10 & D & F \\
P5 & DataVis Designer & 5-7 & M & F \\
P6 & Data Architect & $>$10 & M & M \\
P7 & DataVis Designer & 5-7 & B & F \\
P8 & DataVis Designer & $>$10 & M & M \\
P9 & Senior UX Designer & 5-7 & F \\
P10 & DataVis Journalist/Designer & 5-7 & M & M \\
\hline
\end{tabular}
    \caption{Our 10 participants and their self-reported characteristics: job title, years of experience, highest degree (Bachelors-B, Masters-M and Doctoral-D), and gender. ~\label{tab:participants}}
\end{table}

We first provide an overview of how judgments emerged in the conversations with our participants, specifically highlighting instances where one type of judgment is foregrounded (i.e., brought into focus) and others exist in the background (i.e., related but not focused on). In the next section, we will further describe how these judgment types are complex and layered, with multiple judgment types emerging together.

\subsection{Different Types of Judgments}

\textbf{Framing} judgments (n=25) focus on identifying what elements are salient, often through the introduction of explicit or decisive constraints. In some instances, these judgments were more conceptual and philosophical in nature, such as P1's goal to ``\textit{apply best practices and [\ldots] basically trying to reduce cognitive load and pain points}'' or P3's focus on cognition: ``\textit{And it's about what techniques improve cognition and behavior change, right?}'' For other participants, these judgments represented points they could use to begin their design work, representing an initial problem frame. For P7, this focus was on the client goal and the desired outcome: ``\textit{So the first thing that I need to have clear once I know that I want to accept the project is what is the goal for the client? So what is the main thing that the visualization should do---be able to kind of say that in one or two sentences. What should the people learn and then having a feeling of the data, what variables are in there?}''

\begin{table}
\begin{tabular}{ p{0.20\columnwidth} | p{0.73\columnwidth} }
\hline
\textbf{Judgment Type} & \textbf{Description} \\ 
\hline
Framing & Creating a working area for design activity to occur, often through the introduction of constraints. \\
\hline
Appreciative & Assigning importance to some things, while not to others, without the intervention of hierarchy. \\ 
\hline
Appearance & Making determinations relating to style, nature, character, and experience. \\
\hline
Quality & Making determinations relating to craftspersonship and connoisseurship. \\
\hline
Instrumental & Selecting or reflecting on the influence of tools, methods, or techniques.\\
\hline
Navigational & Identifying and reflecting on a path to achieve a specific design outcome in complex situations. \\
\hline
Compositional & Forming connections among multiple artifacts or concepts, creating a sense of holism or centrality. \\
\hline
Connective & Making connections or bridging design objects to address specific aspects of a design situation. \\ 
\hline
\end{tabular}
    \caption{Operationalized design judgment types used in our analysis (adapted from Nelson and Stolterman \cite{nelson_design_2012} and Gray et al. \cite{gray_judgment_2015}) ~\label{tab:judgments}}
\end{table}





\textbf{Appreciative} judgments (n=43) often occur alongside framing judgments, representing places where certain elements of the design situation were foregrounded, leading other elements to be backgrounded. For some participants, appreciative judgments represented their approach to a design project, such as P6's goal of ``\textit{mak[ing] that easier, simpler, clearer}'' or P9's focus on ''\textit{showing the data that the client wants to show off.}'' Some examples were more precisely grounded in project decisions, such as this example from P1 where they try to foreground different elements of complexity to guide their decisions: ``\textit{There's obviously some seasonal variation; the way it's portrayed now, you can't tell what that seasonal variation is. So I was thinking, okay, do we want to communicate seasonal variation? And if so, let me think about breaking it out this way. Do we want to communicate an overall decline? And if so, let me break it out this way instead.}''





\textbf{Appearance} judgments (n=11) focus on a personal sense of style, character, or experience. While discussing a minimalist style, P5 noted ``\textit{I wouldn't say I'm a minimalist [\ldots] I need to grab the reader's eye. I find that [minimalism] doesn't really work for that. But in terms of embellishments, I always try and keep it very close to the data. So, instead of a color, I might make it a subtle gradient, I might give it a slight drop shadow to emphasize it. Instead of a straight line between two points, I might give it a small curve. So it's like these, I guess smaller additions onto the basics. Things like going beyond the default in a more stylistic way.}'' Appearance judgments do not always relate directly to visual characteristics, however, and may be concerned more with character, form, and material or temporal experience. For example, P2 had a desire to ``\textit{try something cooler}'' if there was time to return to the original data. P7 alluded to a more experiential goal underlying their visual style: ``\textit{I've always tried my best to not have fanciness for fanciness sake [\ldots] I would never be like, let me do a D3 force layout because it bounces around and people think it's fun or something like that. I have used a D3 force layout to make all the nodes burst as people scroll, because that actually got people to scroll. And I wanted them to keep scrolling. So I have some sort of a goal. [\ldots] I think [engaging users] is a really important function.}''






\textbf{Quality} judgments (n=22) focus more on craftspersonship, in comparison to the more subjective qualities of appearance represented earlier. Some participants, such as P1, pointed towards the ``\textit{craft}'' that is often connected with the science of visualization: ``\textit{It's sort of, it's part science. It's part craft. I'd say it's actually more craft.}'' In another example, P1 shared a more extensive example of how they connected their approach to existing design languages to improve the quality of their design, while perhaps reducing the felt creativity of this work: ``\textit{these frameworks [\ldots] in some way homogenize your work so that if you look less creative and more corporate on one hand, on the other hand, I think getting it proves usability because there's familiarity to it.}''



\textbf{Instrumental} judgments (n=48) point towards the influence of tools and their impact on the designer's work  within the design process. While there were a substantial number of examples of specific tools that are commonly used in visualization work, some of the more nuanced examples pointed out the limitations and strengths of various tools and how they might relate to one another. For example, P8 assessed the visual vs. data-focused elements of different tools: ``\textit{In Illustrator it's all about how does this look? [\ldots] sometimes it's somewhat about the data because if I'm in Illustrator, I'm not going to worry about getting everything data--accurate. But if it's a complex visualization or it would just be too hard to execute in illustrator and I go to code first, I literally might write the code to generate the visualization in D3, and in D3 I can export an SVG file and bring that back into Illustrator to then design it up.}'' P10 also shared a more data-specific example, where the form that data takes might encourage certain design paths and discourage others: ``\textit{in a lot of cases, data is shared in the form of excel files and that's not really a format that I can work with. Well so what I usually do is try to get it into R [\ldots]. So that's almost always the first step.}''



\textbf{Navigational} judgments (n=19) point towards the path the designer intends to take to execute their plan as they engage in complex situations. Most of our participant responses related to the navigational elements of their overall design process, where they foregrounded certain tools or sequences of tools they used to undertake their work. For P3, this process started by ``\textit{look[ing] at the material, extract[ing] the data and then start sketching alternatives,}'' and then later ``\textit{pull[ing] it into a visualization platform and start manipulating the data.}'' P10 described their navigational judgments as more situational, noting that they ``\textit{don't really think that there's a typical design process or typical data visualization project,}'' later describing their work as initially more focused on data cleaning. P7 described ``asking questions'' of the data to work through their process ``\textit{when I get the data, the first thing to do is [\ldots] look at the metadata, the attributes [\ldots] and then I start thinking about what might be interesting about that dataset or interesting questions to ask that dataset. [\ldots] And I'll keep trying at it until I get a satisfactory answer or, if I don't, then I note that and I go on to the next answer, and I'll do that until I piece together some sort of a story, some sort of interesting things that I could potentially make into a story.}''



\begin{figure*}
    \centering
    \includegraphics[width=0.90\textwidth]{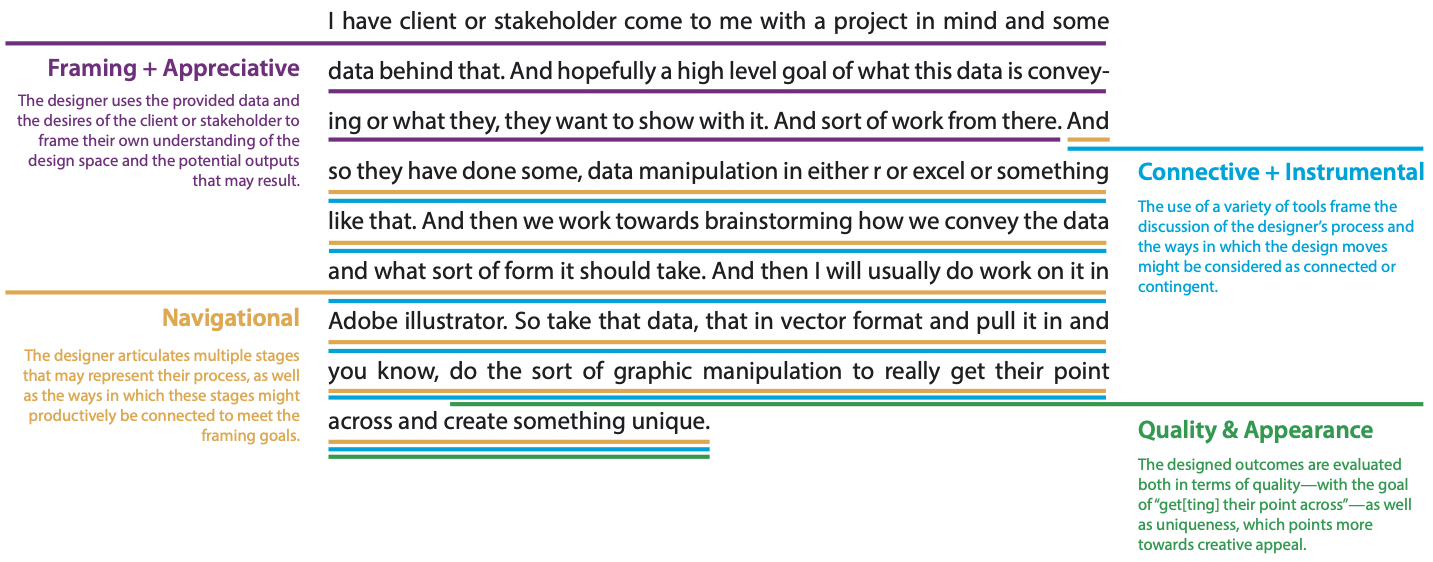}
    \caption{Complex layering of judgments described by P9 as a means of confronting complexity in the design situation.}~\label{fig:layering}
\end{figure*}

\textbf{Compositional} judgments (n=11) identify relationships among design elements, pointing toward an emergent whole as a central, foregrounded concern, with P1 discussing the role of visualization paradigms in helping them ``\textit{think very structurally.}'' P7 pointed towards a narrative focus in their process, describing how they iteratively work to ``\textit{piece together some sort of a story---some sort of interesting things that I could potentially make into a story.}'' In a similar fashion, P10 described creating a ``design layer'' by bringing together multiple concerns to tell a story: ``\textit{First is figuring out what you can do with the data and, within those boundaries, select the best way to visualize it. And then it's more like a design layer that you put over the visualization to make it more compelling and then really letting it tell a story, so to speak, by highlighting different things, playing with colors, playing with fonts. And so yeah, that's the design layer you can play with.}''




\textbf{Connective} judgments (n=19) address the ways in which different elements of design work or designed artifacts are drawn together by the designer. Many of these judgments also pointed towards compositional assemblages as they were operationalized in navigational ways. In one example, P6 describes the ways they connect different judgments relating to the availability of data and its use in creating a visualization: ``\textit{So the broad steps are fairly straight forward, right? So we need to acquire and explore the data. We need to derive any data that we need out of data that doesn't already exist [\ldots]. We need to design the visualization and sort of get approval on that. And then we need to do the work of building the visualization and then promoting it into some environment where it can be consumed.}'' P9 describes a functional assembly of connections that moves their design process: ``\textit{And so they have done some data manipulation in either R or Excel or something like that. And then we work towards brainstorming how we convey the data and what sort of form it should take. And then I will usually do work on it in Adobe Illustrator. So take that data in vector format and pull it in and do the sort of graphic manipulation to really get their point across and create something unique.}''

\subsection{Layering of Judgments}
Our second finding is that judgments typically occur in a complex, layered fashion, where one judgment is often foregrounded while others have influence in the background. Our thematic analysis regularly resulted in multiple codes being applied to statements in an overlapping manner. For instance, consider the excerpt from P9 shown in Figure \ref{fig:layering}. Here we see P9 engaging in multiple, overlapping judgments as a means of confronting the complexity of the design situation. There are framing and appreciative judgments to set the initial design space and potential outputs; connective and instrumental judgments about the tools being used and the ways in which they are connected while moving towards an outcome; navigational judgments that help deal with the complexity of the data in context; and quality and appearance judgments, relating to the creativity and craft of making a point and creating something unique. This complex layering demonstrates how situated, personal judgments are a means of moving through complexity toward a design outcome.


\section{Discussion}
When we asked practitioners to reflect on their design work, they described continuously relying on situated judgments in a layered and complex fashion. This was true across all aspects of their process, including how they start, how they assess their progress and decide what steps to take next, and how they select visual channels and chart types. Rarely did practitioners describe reaching conclusions as a result of rational decision-making processes in which options were weighed and specific steps were followed. This is not to suggest that practitioners are irrational or unprincipled in their work. Many of the practitioners we interviewed are experts in data visualization---they have written influential books, won awards for their work, and are highly active and respected in the practitioner community. Rather, they make judgments that draw from an accumulation of prior experience in combination with a personal design philosophy and various situational factors. Studies in naturalistic decision making show precisely this---that experts draw from familiar, prototypical patterns of experience to make good judgments, relying on a process that mimics intuition more than logical deliberation \cite{klein2017sources,beach2017classical}.

Aside from a brief mention of design judgment \cite{mckenna_design_2014}, the visualization literature has largely promoted rational, model-driven forms of decision-making \cite{munzner_nested_2009,meyer_nested_2015} and moving through a design process \cite{sedlmair_design_2012}. While these popular models may be useful for researchers, our work suggests they are not sufficient for characterizing real-world design practice, especially in relation to the personal and situated ways practitioners confront complex design situations. Decision and process models suggest that design thinking is rational and structured, yet our analysis reveals that practitioners largely do not describe their design activity in this way.  

The emphasis on rational models likely stems from the positivist foundations of field of visualization \cite{meyer_criteria_2020}. Models and other abstract forms of knowledge ignore the complexity and messiness of particular situations in an effort to generalize and be ``objective''. While the objectivity of this kind of knowledge is questionable on its own \cite{polanyi_tacit_1966}, it is also questionable whether rational models have much descriptive or prescriptive value when it comes to decision making in complex environments. Both decision scientists (e.g., \cite{klein2017sources,beach2017classical}) and design scholars (e.g., \cite{Cross_Naughton_Walker_1981,Stolterman2008}) have largely abandoned attempts to model decision making in formal and rationalistic ways. At the very least, our findings indicate that we need more knowledge types---ones that are better able to characterize situated, personal knowledge---if we are to adequately understand and influence visualization design practice. Judgments are one such type of knowledge, which we have shown here to be valuable in characterizing the way designers think and work in practice.

\section{Summary}
Our analysis shows how designers rely on personal and situated forms of knowledge that cannot be not generalized and modeled in the ways that process and decision models tend to be. Our analysis also shows that judgments are complex and layered, rather than taking place in individual and disconnected ways, strengthening the claims made by Gray et al.~\cite{gray_judgment_2015} in another design context. This finding calls into question the adequacy of design models for characterizing design practice---generally speaking, but especially with respect to practitioners designing in non-research settings. 

Our work contributes a new conceptual language and theoretical framing for studying visualization design, particularly in terms of design practice and the ways in which designers face complexity and move through their design process. This contribution can also be valuable for practitioners, as surfacing this way of knowing can help them become aware of their own judgment making and identify means to improve it. Our work has implications for data visualization pedagogy, as an appreciation of the personal and situated nature of design is critical for preparing designers to face the complexities of real-world practice. Future scholarship on data visualization pedagogy and practice would benefit from focusing attention beyond formal, objective knowledge and logical processes of decision making, allowing access to the rich nature of design expertise and the ways in which this expertise is developed over time.



\acknowledgments{
We would like to thank our interview participants and the anonymous reviewers of this paper. This work was supported in part by NSF award 1755957.}

\bibliographystyle{abbrv-doi}

\bibliography{template}

\begin{thebibliography}{10}

\bibitem{beach2017classical}
L.~R. Beach and R.~Lipshitz.
\newblock Why classical decision theory is an inappropriate standard for
  evaluating and aiding most human decision making.
\newblock In G.~Klein, J.~Orasanu, R.~Calderwood, and C.~E. Zsambok, eds., {\em
  Decision Making in Action: Models and Methods}, pp. 21--35. Ablex Publishing,
  1993.

\bibitem{bigelow_reflections_2014}
A.~Bigelow, S.~Drucker, D.~Fisher, and M.~Meyer.
\newblock Reflections on how designers design with data.
\newblock In {\em Proceedings of the 2014 {International} {Working}
  {Conference} on {Advanced} {Visual} {Interfaces} ({AVI} '14)}, pp. 17--24.
  ACM Press, Como, Italy, 2014. doi: {{%
10\hspace{.1pt}\discretionary{.}{%
}{.}\hspace{.4pt}1145\discretionary{/}{%
}{/}2598153\hspace{.1pt}\discretionary{.}{%
}{.}\hspace{.4pt}2598175}}


\bibitem{bigelow2016iterating}
A.~Bigelow, S.~Drucker, D.~Fisher, and M.~Meyer.
\newblock Iterating between tools to create and edit visualizations.
\newblock {\em IEEE Transactions on Visualization and Computer Graphics},
  23(1):481--490, 2016.

\bibitem{boling_core_2017}
E.~Boling, H.~Alangari, I.~M. Hajdu, M.~Guo, K.~Gyabak, Z.~Khlaif,
  R.~Kizilboga, K.~Tomita, M.~Alsaif, A.~Lachheb, H.~Bae, F.~Ergulec, M.~Zhu,
  M.~Basdogan, C.~Buggs, A.~Sari, and R.~Techawitthayachinda.
\newblock Core {Judgments} of {Instructional} {Designers} in {Practice}.
\newblock {\em Performance Improvement Quarterly}, 30(3):199--219, 2017. doi:
  {{%
10\hspace{.1pt}\discretionary{.}{%
}{.}\hspace{.4pt}1002\discretionary{/}{%
}{/}piq\hspace{.1pt}\discretionary{.}{%
}{.}\hspace{.4pt}21250}}


\bibitem{braun2006using}
V.~Braun and V.~Clarke.
\newblock Using thematic analysis in psychology.
\newblock {\em Qualitative Research in Psychology}, 3(2):77--101, 2006.

\bibitem{chi_taxonomy_2000}
E.~Chi.
\newblock A taxonomy of visualization techniques using the data state reference
  model.
\newblock pp. 69--75. IEEE Symposium on Information Visualization 2000, 2000.
  doi: {{%
10\hspace{.1pt}\discretionary{.}{%
}{.}\hspace{.4pt}1109\discretionary{/}{%
}{/}INFVIS\hspace{.1pt}\discretionary{.}{%
}{.}\hspace{.4pt}2000\hspace{.1pt}\discretionary{.}{%
}{.}\hspace{.4pt}885092}}


\bibitem{Cross_1982}
N.~Cross.
\newblock Designerly ways of knowing.
\newblock {\em Design Studies}, 3(4):7, 1982.

\bibitem{Cross_Naughton_Walker_1981}
N.~Cross, J.~Naughton, and D.~Walker.
\newblock Design method and scientific method.
\newblock {\em Design Studies}, 2(4):195–201, 1981.

\bibitem{Demiral-Uzan2015-al}
M.~Demiral-Uzan.
\newblock Instructional design students' design judgment in action.
\newblock {\em Performance Improvement Quarterly}, 28(3):7--23, Oct. 2015. doi:
  {{%
10\hspace{.1pt}\discretionary{.}{%
}{.}\hspace{.4pt}1002\discretionary{/}{%
}{/}piq\hspace{.1pt}\discretionary{.}{%
}{.}\hspace{.4pt}21195}}


\bibitem{dunne_professional_1999}
J.~Dunne.
\newblock Professional judgment and the predicaments of practice.
\newblock {\em European Journal of Marketing}, 33(7/8):707--720, 1999. doi: {{%
10\hspace{.1pt}\discretionary{.}{%
}{.}\hspace{.4pt}1108\discretionary{/}{%
}{/}03090569910274339}}


\bibitem{goodman_understanding_2011}
E.~Goodman, E.~Stolterman, and R.~Wakkary.
\newblock Understanding interaction design practices.
\newblock In {\em Proceedings of the 2011 SIGCHI Conference on {Human} Factors
  in Computing Systems ({CHI} '11)}, p. 1061. Vancouver, BC, Canada, 2011. doi:
  {{%
10\hspace{.1pt}\discretionary{.}{%
}{.}\hspace{.4pt}1145\discretionary{/}{%
}{/}1978942\hspace{.1pt}\discretionary{.}{%
}{.}\hspace{.4pt}1979100}}


\bibitem{gray_judgment_2015}
C.~M. Gray, C.~Dagli, M.~Demiral-Uzan, F.~Ergulec, V.~Tan, A.~A. Altuwaijri,
  K.~Gyabak, M.~Hilligoss, R.~Kizilboga, K.~Tomita, and E.~Boling.
\newblock Judgment and {Instructional} {Design}: {How} {ID} {Practitioners}
  {Work} {In} {Practice}.
\newblock {\em Performance Improvement Quarterly}, 28(3):25--49, 2015. doi: {{%
10\hspace{.1pt}\discretionary{.}{%
}{.}\hspace{.4pt}1002\discretionary{/}{%
}{/}piq\hspace{.1pt}\discretionary{.}{%
}{.}\hspace{.4pt}21198}}


\bibitem{hoffswell2020techniques}
J.~Hoffswell, W.~Li, and Z.~Liu.
\newblock Techniques for flexible responsive visualization design.
\newblock In {\em Proceedings of the 2020 SIGCHI Conference on Human Factors in
  Computing Systems (CHI '20)}, pp. 1--13, 2020.

\bibitem{klein2017sources}
G.~A. Klein.
\newblock {\em Sources of power: How people make decisions}.
\newblock MIT press, 2 ed., 2017.

\bibitem{kuutti_turn_2014}
K.~Kuutti and L.~J. Bannon.
\newblock The turn to practice in {HCI}: towards a research agenda.
\newblock In {\em Proceedings of the 2014 SIGCHI Conference on {Human} Factors
  in Computing Systems ({CHI} '14)}, pp. 3543--3552. ACM Press, Toronto,
  Canada, 2014. doi: {{%
10\hspace{.1pt}\discretionary{.}{%
}{.}\hspace{.4pt}1145\discretionary{/}{%
}{/}2556288\hspace{.1pt}\discretionary{.}{%
}{.}\hspace{.4pt}2557111}}


\bibitem{lawson2006designers}
B.~Lawson.
\newblock {\em How designers think: The design process demystified}.
\newblock Routledge, 2006.

\bibitem{McCurdy2016}
N.~McCurdy, J.~Dykes, and M.~Meyer.
\newblock Action {Design} {Research} and {Visualization} {Design}.
\newblock In {\em Proceedings of the 6th {Biannual} {Workshop} on evaluation
  and {BEyond} - {methodoLogIcal} approaches for {Visualization} ({BELIV})},
  pp. 10--18. ACM Press, New York, New York, USA, 2016. doi: {{%
10\hspace{.1pt}\discretionary{.}{%
}{.}\hspace{.4pt}1145\discretionary{/}{%
}{/}2993901\hspace{.1pt}\discretionary{.}{%
}{.}\hspace{.4pt}2993916}}


\bibitem{mckenna_design_2014}
S.~McKenna, D.~Mazur, J.~Agutter, and M.~Meyer.
\newblock Design {Activity} {Framework} for {Visualization} {Design}.
\newblock {\em IEEE Transactions on Visualization and Computer Graphics},
  20(12):2191--2200, 2014. doi: {{%
10\hspace{.1pt}\discretionary{.}{%
}{.}\hspace{.4pt}1109\discretionary{/}{%
}{/}TVCG\hspace{.1pt}\discretionary{.}{%
}{.}\hspace{.4pt}2014\hspace{.1pt}\discretionary{.}{%
}{.}\hspace{.4pt}2346331}}


\bibitem{mendez2017bottom}
G.~G. M{\'e}ndez, U.~Hinrichs, and M.~A. Nacenta.
\newblock Bottom-up vs. top-down: trade-offs in efficiency, understanding,
  freedom and creativity with infovis tools.
\newblock In {\em Proceedings of the 2017 SIGCHI Conference on Human Factors in
  Computing Systems (CHI '17)}, pp. 841--852, 2017.

\bibitem{meyer_criteria_2020}
M.~Meyer and J.~Dykes.
\newblock Criteria for {Rigor} in {Visualization} {Design} {Study}.
\newblock {\em IEEE Transactions on Visualization and Computer Graphics},
  26(1):87--97, 2020. doi: {{%
10\hspace{.1pt}\discretionary{.}{%
}{.}\hspace{.4pt}1109\discretionary{/}{%
}{/}TVCG\hspace{.1pt}\discretionary{.}{%
}{.}\hspace{.4pt}2019\hspace{.1pt}\discretionary{.}{%
}{.}\hspace{.4pt}2934539}}


\bibitem{meyer_nested_2015}
M.~Meyer, M.~Sedlmair, P.~S. Quinan, and T.~Munzner.
\newblock The nested blocks and guidelines model.
\newblock {\em Information Visualization}, 14(3):234--249, 2015. doi: {{%
10\hspace{.1pt}\discretionary{.}{%
}{.}\hspace{.4pt}1177\discretionary{/}{%
}{/}1473871613510429}}


\bibitem{munzner_nested_2009}
T.~Munzner.
\newblock A nested model for visualization design and validation.
\newblock {\em {IEEE} {Transactions} on {Visualization} and {Computer}
  {Graphics}}, 15:921--928, 2009. doi: {{%
10\hspace{.1pt}\discretionary{.}{%
}{.}\hspace{.4pt}1109\discretionary{/}{%
}{/}TVCG\hspace{.1pt}\discretionary{.}{%
}{.}\hspace{.4pt}2009\hspace{.1pt}\discretionary{.}{%
}{.}\hspace{.4pt}111}}


\bibitem{nelson_design_2012}
H.~G. Nelson and E.~Stolterman.
\newblock {\em The {Design} {Way}: {Intentional} {Change} in an {Unpredictable}
  {World}}.
\newblock The MIT Press, 2 ed., 2012.

\bibitem{parsons_what_2020}
P.~Parsons, A.~Baigelenov, Y.-H. Hung, and C.~Schrank.
\newblock What {Design} {Methods} do {DataVis} {Practitioners} {Know} and
  {Use}?
\newblock In {\em Extended {Abstracts} of the 2020 {CHI} {Conference} on
  {Human} {Factors} in {Computing} {Systems}}, pp. 1--8. ACM, Honolulu HI USA,
  2020. doi: {{%
10\hspace{.1pt}\discretionary{.}{%
}{.}\hspace{.4pt}1145\discretionary{/}{%
}{/}3334480\hspace{.1pt}\discretionary{.}{%
}{.}\hspace{.4pt}3383048}}


\bibitem{Parsons_chartjunk_2020}
P.~Parsons and P.~Shukla.
\newblock Data visualization practitioners' perspectives on chartjunk.
\newblock In {\em Proceedings of the 2020 IEEE Visualization Conference (VIS),
  short papers}, 2020.

\bibitem{polanyi_tacit_1966}
M.~Polanyi.
\newblock {\em The {Tacit} {Dimension}}.
\newblock University of Chicago Press, 1966.

\bibitem{ren2018reflecting}
D.~Ren, B.~Lee, M.~Brehmer, and N.~H. Riche.
\newblock Reflecting on the evaluation of visualization authoring systems:
  Position paper.
\newblock In {\em 2018 IEEE Evaluation and Beyond-Methodological Approaches for
  Visualization (BELIV)}, pp. 86--92. IEEE, 2018.

\bibitem{satyanarayan2019critical}
A.~Satyanarayan, B.~Lee, D.~Ren, J.~Heer, J.~Stasko, J.~Thompson, M.~Brehmer,
  and Z.~Liu.
\newblock Critical reflections on visualization authoring systems.
\newblock {\em IEEE Transactions on Visualization and Computer Graphics},
  26(1):461--471, 2019.

\bibitem{sedlmair_design_2012}
M.~Sedlmair, M.~Meyer, and T.~Munzner.
\newblock Design {Study} {Methodology}: {Reflections} from the {Trenches} and
  the {Stacks}.
\newblock {\em IEEE Transactions on Visualization and Computer Graphics},
  18(12):2431--2440, 2012. doi: {{%
10\hspace{.1pt}\discretionary{.}{%
}{.}\hspace{.4pt}1109\discretionary{/}{%
}{/}TVCG\hspace{.1pt}\discretionary{.}{%
}{.}\hspace{.4pt}2012\hspace{.1pt}\discretionary{.}{%
}{.}\hspace{.4pt}213}}


\bibitem{Shotter2014-ev}
J.~Shotter and H.~Tsoukas.
\newblock Performing phronesis: On the way to engaged judgment.
\newblock {\em Management Learning}, July 2014.

\bibitem{Stolterman2008}
E.~Stolterman.
\newblock The {Nature} of {Design} {Practice} and {Implications} for
  {Interaction} {Design} {Research}.
\newblock {\em International Journal of Design}, 2(1):55--65, 2008. doi: {{%
10\hspace{.1pt}\discretionary{.}{%
}{.}\hspace{.4pt}1016\discretionary{/}{%
}{/}j\hspace{.1pt}\discretionary{.}{%
}{.}\hspace{.4pt}phymed\hspace{.1pt}\discretionary{.}{%
}{.}\hspace{.4pt}2007\hspace{.1pt}\discretionary{.}{%
}{.}\hspace{.4pt}09\hspace{.1pt}\discretionary{.}{%
}{.}\hspace{.4pt}005}}


\bibitem{Vickers1984-gc}
S.~G. Vickers.
\newblock {Judgment}.
\newblock In {\em The Vickers Papers}, pp. 230--245. Harper \& Row, London,
  Jan. 1984.

\bibitem{walny2019data}
J.~Walny, C.~Frisson, M.~West, D.~Kosminsky, S.~Knudsen, S.~Carpendale, and
  W.~Willett.
\newblock Data changes everything: Challenges and opportunities in data
  visualization design handoff.
\newblock {\em IEEE Transactions on Visualization and Computer Graphics},
  26(1):12--22, 2019.

\end{thebibliography}
\end{document}